\documentclass[preprint2]{aastex} 

\newcommand{\eg}{{\it e.g.,\ }}

\newcommand{\etal}{{\it et al.\ }}

\newcommand{\degrees}{$^\circ$}

\newcount\notenumber
\newcount\eqnumber
\newcount\fignumber
\newbox\abstr
\newbox\figca

\def\spose#1{\hbox to 0pt{#1\hss}}    
\def\ltsim{\mathrel{\spose{\lower 3pt\hbox{$\mathchar"218$}}
     \raise 2.0pt\hbox{$\mathchar"13C$}}}
\def\gtsim{\mathrel{\spose{\lower 3pt\hbox{$\mathchar"218$}}
     \raise 2.0pt\hbox{$\mathchar"13E$}}}
     
\def\gg{\hbox{$>$\hskip-4pt $>$}}
\begin{document} 
\title{\bf Some Implications of the Anisotropic Distribution of 
Satellite Galaxies}

\author{Dennis Zaritsky\altaffilmark{1} and Anthony H. Gonzalez}
\affil{UCO/Lick Observatories and Board of Astronomy and
Astrophysics,} 
\affil{Univ. of California at Santa Cruz, Santa Cruz, CA, 95064}
\altaffiltext{1}{Current address: Steward Observatory, Univ. of Arizona, Tucson, AZ, 85721}
\email{dzaritsky@as.arizona.edu, anthonyg@ucolick.org}

\begin{abstract}
We examine a possible connection between 
the anisotropic distribution
of satellite galaxies around giant spiral galaxies
and the evolution of satellite systems. The observed
polar anisotropy (Zaritsky \etal 1997) is either
imprinted by initial conditions or
develops from an initially symmetric distribution.
We attempt to discriminate between
these two possibilities by exploring the implications of the latter
one. From the observed distribution of
satellite galaxies relative to the primary galaxy's disk, we 
derive constrains on the orbital inclinations of the current
satellite population. Using this derived inclination limit and
assuming that the initial population had no preferred orbital
inclination, we estimate the size of the hypothesized original 
population. We find that our best-fit models imply a 
population of destroyed (or inhibited) 
satellites whose combined luminosity (assuming the same
$M/L$ as for the observed satellites)
is between 18\% and 103\% of the current disk luminosity. 
\end{abstract}


\section{Introduction}

Disk galaxies presumably form from protogalactic clouds that 
consist of at least several sub-galaxy 
aggregates, some fraction of which eventually merge to form the dominant
galaxy.
How efficient and complete is this hierarchical process?
Quantitative answers to this type of question have been solely in the
domain of simulations (cf. Navarro and Steinmetz 1997
; Klypin \etal 1999 (hereafter K99); Moore \etal 1999 (hereafter M99)).
Those
simulations are in turn constrained by observations
of the properties of present day galaxies (such as the local
Tully-Fisher relation),
which are the result of a variety of complicated physical 
processes, 
and of the properties of high redshift galaxies, which are difficult to
quantify and influenced by selection biases. A reasonable 
goal is to find a more direct
link to the process of hierarchical formation. 

In hierarchical formation scenarios, small mass objects 
generally collapse prior to large ones and become
the building blocks of larger objects.
Two types of objects that currently surround giant galaxies may qualify 
as possible hierarchical building blocks: 
globular clusters and satellite galaxies. Most studies of
the dynamical evolution of either globular clusters or satellite
galaxies
begin with an assumed initial distribution of such companions and
focus on the subsequent dynamical evolution of the system. 
For example, Aguilar,
Hut, \& Ostriker (1988) calculate the rate of destruction of globular
clusters to assess whether the galactic spheroid was built from
globular clusters, and Ostriker \& Tremaine (1975) investigate 
the luminosity evolution of the primary galaxy due to infalling satellites.
A difficulty with this approach 
is that the results depend sensitively on the unknown
characteristics of the initial population and that we have no empirical
means of determining whether the observed population represents
a small or large fraction of the initial population. 
High-resolution numerical simulations are beginning to produce
populations of low-mass companions around giant galaxies in 
a self-consistent cosmological framework (K99 and M99), but they
produce far more satellite galaxies than observed.

The observed distribution of satellite galaxies of spiral primaries 
out to 500 kpc is
asymmetric and elongated along the disk minor axis (Odewahn 1989;
Zaritsky et al. 1997,
hereafter ZSFW, and $H_0 = 75$ km/s/Mpc assumed throughout). 
This elongation is an extension
of the Holmberg effect (the preferred polar orientation of
satellites interior to $r \sim 50$ kpc;
Holmberg 1969) to larger radii --- although the physical 
causes of the two observational results may differ and the polar elongation
is not evident at intermediate radii (50 to 200 kpc). Both the
inner and outer satellite results are statistical
because orbits of individual satellites are unknown
(although the ZSFW result is for kinematically
confirmed satellite galaxies). 
In the one galaxy for which individual satellite
orbits are known, the Milky
Way, there is evidence that the orbits are preferentially 
polar from the alignment
of satellites on the sky (Kunkel \& Demers 1976; Lynden-Bell 1982), 
the orientation of the Magellanic Stream (Mathewson, Cleary, and
Murray 1974),
the three-dimensional distribution of satellites (Majewski 1994;
Hartwick 1996), and
their space velocities (Scholz \& Irwin 1994).
Finally, Grebel, Kolatt, \& Brandner
(1999) find tentative evidence for a statistical excess of
M 31 satellites along a polar orbit.
Hence, preferentially polar satellite orbits
may be common.

The connection between disks and satellite orbits
is either imprinted in the initial conditions
or is the result of dynamical phenomena during the formation and
subsequent evolution of the galaxy. 
ZSFW argue that the orbital decay time due to dynamical friction at radii
larger than 200 kpc excludes dynamical friction as the dominant
mechanism. For example, models of the
effect of dynamical friction on the orbit of the Large Magellanic Cloud 
(Tremaine 1976) suggest that
the perigalacticon distance has decreased by a factor of $\sim$ 3 in 10 Gyr. 
Therefore, satellites similar to the Large Magellanic Cloud
that began at perigalacticon radii $>$ 200 kpc will
not have merged with their parent galaxies.
Quinn \& Goodman (1986), in their investigation of the 
Holmberg effect, found that dynamical friction
could not even account for the asymmetry inside 50 kpc.

If we wish to invoke a dynamical process for the origin of the
anisotropy, 
we must hypothesize that the missing satellites experienced
a catastrophic event that either destroyed the satellite
or inhibited the formation of stars (such as the removal
of gas from the proto-satellite).
Such an event is most likely to occur as satellites make
a pericentric pass near the giant, and so could
only affect satellites that have made at least one
pericenter passage. Using the standard orbital 
equations of the timing argument (Kahn \& Woltjer
1959), a halo mass of 1.5$\times 10^{12} M_\odot$ 
(the 90\% confidence lower mass limit for the mass enclosed at 200 kpc
for this sample of primary galaxies:
Zaritsky \& White 1994), and 
$t_0 = 15$ Gyr for the age of the Universe, we find that
any satellite on a radial orbit at a current distance $<$ 530 kpc
has made at least one pericenter passage.
If (1) satellite orbits are highly radial (as found in 
recent simulations: K99 and M99)
and (2) satellites on planar orbits either preferentially
lose a greater amount of orbital energy, have their gas removed, 
or are disrupted near pericenter,
then anisotropy in the satellite population might extend
to radii as large as 500 kpc. The currently available simulations (K99 and 
M99)
do not show such satellite destruction, but these models do not include
gas and the survivability of satellites is
sensitive to the particulars of the simulation, such as
details of the power spectrum (M99).
Whether satellite destruction is more common for satellites on
planar orbit and whether the mechanism is sufficiently severe
to induce the asymmetry has not been demonstrated 
and must be investigated further.
To continue our exploration of the 
evolution hypothesis, we postulate that such a mechanism
does exist and follow the argument to its conclusion.
 
Regardless of the exact dynamical model that may lead to 
the anisotropy in the evolution conjecture, we can use the anisotropy
to estimate the toll that the process has
exacted on the satellite
population. We do this by (1) constraining the 
range of satellite orbital inclinations allowed by 
the ZSFW sample and (2)
estimating the size of the initial satellite
population by assuming that it was initially spherically symmetric.
Is the inferred missing population a significant component
of the galaxies or are the results so implausible that
they enable us to exclude the evolution conjecture? 
The methods used to constrain the orbital inclinations are 
discussed in \S2. After determining the number of ``missing'' satellites,
we assess whether the destroyed 
satellites constitute a significant fraction of the mass of the primary
galaxy. The results and implications are discussed in \S3.

\section{Determining the Orbital Inclination Limit}

Because projection effects and the wide range of viewing angles
partially mask
any underlying asymmetric satellite distribution, an observed asymmetric 
distribution implies a more strongly asymmetric underlying distribution.
To determine the degree of polar alignment necessary to
reproduce the observed distribution, we determine the 
orbital inclination limit in three different ways. In all three
ways we presume that there is a single lower inclination limit for
satellite orbits.  

Our first approach is adopted from Quinn and Goodman's (1986)
treatment. For assumed circular orbits and a power-law radial
density profile (parameterized by $\rho \propto R^\beta$), 
they derive an analytic expression for the 
surface density as a function of angle from the disk plane.
In Figure 1, we plot the number of satellites as a function
of angle from the plane, $\theta$, and compare the results
from Quinn and Goodman's calculation for an orbital inclination limit
of 45$^\circ$ and $\beta = 1.8$ (as measured for
satellite galaxies : Lake \& Tremaine 1980, Zaritsky \etal 1993,
Lorrimer \etal 1994)). 
For comparison, we plot the number of satellites 
vs. $\theta$ for satellites at $r > 200$ kpc (for which the
anisotropy appears stronger). This comparison suggests that 
an orbital inclination limit of 45$^\circ$ (we define
the inclination limit to be 
measured from the pole) is appropriate for the full
sample and that this limit is tighter
for the outer satellites. 

Quinn \& Goodwin's calculation is independent
of the assumption of circular orbits, as long as the orbits
are not closed and one time averages as the apsides of
eccentric orbits precess. However, the satellites at large
radii in our sample
have not completed many orbits, the apsides have not precessed, and so
this assumption may be inadequate.
A second possible shortcoming of the calculation
is that interlopers, apparent satellites
that are not physically associated with the system, are not included.
The estimated fraction of interlopers for this sample is between
10 and 15\% (Zaritsky 1992).

We proceed by examining models with
Keplerian orbits that include interlopers.
The satellite orbits are taken from a family of fixed eccentricity
orbits for any single model 
(although we explore a range of eccentricity values across all models).
The orbital energy is drawn from the power-law distribution 
given by Bahcall and Tremaine (1981)
$$P(E) = \cases{(3-s)E^{s-4}/E_0^{s-3},&if $E > E_0$;\cr
0,&if $E \le E_0,$\cr}$$
where $s < 3$ and $E = (GM/r)-(v^2/2)$. This choice of $P(E)$
generates
a number density profile of test particles that is proportional 
to $r^{-s}$ for $E > E_0$. We choose
$s=1.8$ and $E_0 = 0.0065$ to match the observed mean projected
separation, $\langle r_p
\rangle$ ($\sim 200$ kpc), and the radial number density
profile (Lake \& Tremaine 1980; 
Zaritsky \etal 1993; Lorrimer \etal 1994). The mass ratio
between
the satellite and primary is chosen to be 1:20 (comparable
to the mean observed ratio assuming equal M/L's for primary and
satellite which is $\sim 1:13$). 
The mean anomaly
(orbital phase) is selected uniformly from $(0,2\pi]$. The satellite orbits
are then randomly oriented using the Euler angle convention and the
known distribution of primary disk inclinations for the ZSFW primaries. 
Simulated orbits
are accepted only if the angle between the orbital major axis and the primary
disk's rotation axis is $\le \theta_l$, where $\theta_l$
is the orbital inclination limit. In exploring
the models we vary $s$, $e$, $\theta_l$, and $f_{INT}$, where
$f_{INT}$ is the fraction of the sample that consists of interlopers.
From observations, we limit $s$ to between 1.5 and 2 (Lorrimer \etal 1994).
On the basis of ZW's infall simulations we limit $e$ to between 0.5
and 0.9, with a preferred value of 0.7. Finally, from various
arguments (Zaritsky 1992), our preferred value of $f_{INT}$ is 
0.1, but we also explore $f_{INT} = 0.05$ and 0.15.
For each model, we generate 10,000 artificial satellites. 

We compare the results of these simulations to four subsamples
of the ZSFW satellite galaxies. 
Sample 1 includes all of the satellites of all of the primaries in the
ZSFW sample. Sample 2 includes only those satellites beyond 
$r_p = $ 300 kpc, and so is limited to the radial range where a
strong azimuthal asymmetry is evident (Figure 2). 
Sample 3 includes all of the 
satellites of the primaries with disk inclination angles $>$
45$^\circ$. This sample is less affected by projection
and confusion between polar orbits and those in the disk plane.
Sample 4 includes only those satellites beyond 
$r_p = $ 200 kpc that are associated with primaries with disk
inclination angles $>$  45$^\circ$. The inner radial limit for Sample 4
is decreased from 300 to 200 kpc relative to Sample 2
because the asymmetry is evident in this
sample down to $r_P \sim 200$ kpc and the number of satellites beyond 
$r_P = $ 300 kpc is smaller (this information is summarized
in Table 1). To compare the
simulations with the data, we calculate the two-sided KS 
statistic\footnote{This statistic is better suited to the analysis of 
data with 
no natural minimum or maximum (as is the case for position angles)
than the standard KS test (Press \etal 1992).} for the distributions 
of satellite position angles relative to the disk major axis. The
results for the best fit $\theta_l$ and 90\% confidence interval 
are presented for various models and data samples in Column (6) in
Table 2.
Again, we conclude that the most likely value of the orbital inclination
limit for the full sample lies around 45$^\circ$ and that this
limit for the outer sample is smaller (about
20$^\circ$). However, the 
uncertainties on the derived $\theta_l$ are large.

As discussed by Zaritsky and White (1994), the full 
description of the dynamics of 
satellite systems requires a model of the growth of the primary
galaxy's halo with time and the evolution of the satellite population 
within that halo. In particular, the assumption that satellites are 
currently found at a random phase along an orbit (which is
necessary for the Keplerian models) is suspect for 
satellites at large radii, where the orbital period is $\sim$ 
Hubble time. Therefore,
we proceed to test the results from the Keplerian models (for which
parameter space is easily explored) with the results of the
spherical-infall halo simulations used by Zaritsky \& White (1994)
to measure the mass of galaxy halos. Using the simulation
that best matches their best fit parameters ($\Omega_0 = 0.3$),
we have derived preferred orbital inclination limits for
the four satellite subsets. We present those results in Column (7)
of Table 2. The best fit values are indistinguishable from
those derived using the Keplerian models, but the 90\% confidence
ranges vary. In particular, we find that the entire range
of $\theta_l$ is allowed when the satellite sample includes satellites
at all radii, and that none of the range is within the 90\% confidence
limit when only the outer satellites are considered (indicated by 
$[$---,---$]$). 

We conclude that all three analysis techniques indicate similar
best fit limits ($\sim 15^\circ$ to 60$^\circ$), 
but that strong (\eg 90\% confidence) statistical conclusions
cannot yet be reached. Because of the agreement among the
various methods used to determine $\theta_l$, 
the current principal limitation does not
appear to lie in the details of the models, but rather with 
the sample size. 

\section{Discussion}

For all samples and all model parameters within our specified ranges, 
the best fit values of $\theta_l$ indicate that the orbits are 
preferentially polar. The best fit $\theta_l$ for Sample 1 and 
our reference model 
($s = 1.8$, $e = 0.7$, and $f_{INT} = 0.1$) indicates that all
of the satellites are on orbits that are inclined at least 38$^\circ$ to
the disk plane ($\theta_l = 52^\circ$). Over the
range of radii where the asymmetry is most pronounced ($> 200$ kpc)
for systems with primaries highly inclined to the line-of-sight
($>45^\circ$), the best-fit 
solutions indicate that the orbits are confined to within 
20$^\circ$ of the pole (for either Keplerian or Infall models) and that 
the orbits are confined to within $\sim 60^\circ$
of the pole with greater than 90\% confidence. 

Before drawing conclusions from these results, 
we discuss the sensitivity of the models to
various parameters. Our reference model is defined to have $e=0.7$,
$s= 1.8$,  and $f_{INT} = 0.1$ (the results from this
 model, as applied to Sample 4, are presented as Model 2 in Table 2).
We test all of these choices with the Keplerian models. First, 
we vary $e$ between 0.5 and 0.9 (the 90\% confidence limits derived
by ZW; Models 1 and 3). 
This parameter sometimes has a noticeable effect on $\theta_{l}$, so
we present results for all three eccentricities. 
Second, we vary $s$ between 1.5 and 2.0 
(Models 4 through 9 in Table 2). The results are nearly insensitive to $s$. 
Third, we vary
the interloper fraction between 5 and 15\% (Models 10 through 15). 
As with $s$, changing
this parameter has a minimal effect on $\theta_{l}$. 
We also present results derived using other subsamples (for the
standard parameter choices and $0.5 \le e \le 0.9$; Models 16 through
24). 

We now determine the number of ``missing'' satellites
implied by a particular $\theta_l$.
From the range of allowed orbital inclinations, we calculate the fraction
of all allowed orbits that are represented in the current sample.
This fraction is equivalent to the fraction of the volume of
a sphere that lies within $\theta_l$ of the pole, which equals
$1 - \cos \theta_l$.
For an opening angle of 30\degrees, the 
volume within the allowed cone is 13.4\% of that within the sphere. 
If the ``original'' population 
of satellite orbits uniformly filled the sphere and our
value for $\theta_l$ is $30^\circ$, then the current population is
only 13.4\% of the original population, or originally there were
7.5 times as many satellites as there are in the observed sample. 

The interpretation of $\theta_l$ and the ``missing''
satellite population is 
complicated by the apparent change in the magnitude
of the polar asymmetry at different radii. 
The data in Figure 2 suggest that in addition to the 
asymmetry at $r_P > 300$ kpc, there may be a slight excess of planar
satellites at $\sim$ 180 kpc and a slight excess of polar satellites
once again at small radii. 
Although our simulations illustrate that projection effects result
in less apparent asymmetry at small radii even in a model
where the limit on orbital inclination is the same at all radii (see
Figure 2), we do not reproduce the observed dip in polar angle
at $\sim 180$ kpc. However, 
this apparent disagreement is not statistically significant (
significance $\sim 1 \sigma$) because of the small number of satellites 
in each radial bin. The gradual radial increase
in the number of satellites ($\propto r^{0.2}$) and
the lack of a strongly planar asymmetry between 0 and 200 kpc suggest that 
a large number of outer, planar satellites cannot be hidden as
satellites at smaller radii,
{\it unless} there is a destruction of a comparable number of 
inner planar satellites to compensate. Therefore, we can calculate
the number of ``missing'' outer satellites and use that quantity as an
estimate of the total number of ``missing'' satellites. 

The dependence of the polar asymmetry on radius leads to 
the result that even an orbital inclination limit of 90\degrees\ 
is allowed within the 90\% confidence limit
when satellites at all radii are included (Samples 1 and 3).
The lack of a strong polar signature in the complete
sample weakens the claim of polar alignment 
and argues for larger samples to resolve the issue. However, we
remind the reader that the polar alignment at large radius is
highly significant (ZSFW) and that polar alignements have also been observed
both in the satellite system of our galaxy (cf. Hartwick 1996) 
and in the Magellanic Irr satellites of other galaxies
(Odewahn 1989).

We now estimate the number of satellites destroyed, inhibited, or accreted
by field giant spiral galaxies.
If $\theta_l = 18^\circ$ (best fit for Keplerian Model 2)
for the outer satellite population, a large population of
corresponding disk plane satellites (19 times the
current number of satellites beyond $r_p = 200$ kpc) 
are missing. For the average primary
galaxy in our sample, 
this estimate implies a loss of 13.4 satellites of comparable
luminosity as the satellites in our sample (for the $i >
45^\circ$ sample).
However, due to the small sample size we cannot exclude the possibility
with greater than 90\% confidence that the missing population is only 
comparable
in size to the current population (0.7 satellites/primary at $r_p >
200$ kpc). The latter conclusion obviously places less
stringent constraints on the possible current status of the missing 
satellites. 

Over the suite of models and samples, we infer a wide range in the
number of missing satellites.
If we adopt our best fits for $\theta_l$ across all samples for
our reference model, then 
the number of missing satellites inferred per galaxy ranges
from 2.4 to 13.4.
These values for the population of missing satellites
are consistent with the number of ``extra'' 
satellites (with velocity dispersion $\gtsim 30$ km sec$^{-1}$)
present in recent simulations (K99 and M99).
The average luminosity of a primary in our sample is 13 times
the average luminosity of a satellite (this calculation includes
a completeness correction factor of 1.4 due to satellites that
may have been missed in our spectroscopic survey as
derived assuming a Schechter luminosity function with a faint-end
slope $\alpha = -1.5$). Therefore, if the 
primaries have indeed accreted between 2.4 and 13.4 satellites, a
significant fraction of their luminosity (18 to 103\%), 
and possibly their mass if the M/L's are comparable, 
comes from these satellites. 

The 
best fit $\theta_l$ value for all of the satellites of all the primaries 
(52$^\circ$) implies that we are missing 2.7
satellites per primary (or about 20\% of the disk luminosity). 
If we assume that these satellites have been accreted, we can compare
this value to the estimates of the satellite accretion rate from
other studies.
An extrapolation of the local
accretion rate (Zaritsky \& Rix 1997) predicts that 1 to
3 large satellites (7 to 21\% of the current luminosity)
are accreted over the lifetime of the galaxy. 
The actual number of satellites accreted over the lifetime
of the galaxy is likely to be 
larger than the extrapolation of this estimate
because the interaction rate is expected to 
increase with redshift. Within the large uncertainties 
in both approaches, the inferred satellite accretion rates 
are consistent and imply that satellite material may contribute 
significantly to the luminosity of the central galaxy.
Interestingly, our investigation
did not lead to predictions of satellite populations
that had $L_{Total} \gg L_{Disk}$, which would be implausible,
or that had $L_{Total} \ll L_{Disk}$, which would have made this discussion
academic.

We conclude that the evolution conjecture for the polar asymmetry
has the following intriguing implications: (1) it enables an
estimate of the size of the
original satellite population, (2) the inferred population
of ``missing'' satellites would have a luminosity of order that of the
disk, and (3) the inferred number of missing satellites is 
consistent with the excess number of satellites produced by
the most recent numerical simulations of galaxy formation (for
satellites
with velocity dispersions $\gtsim$ 30  km sec$^{-1}$). The principal
difficulty with the evolution conjecture remains the unidentified
physical mechanism necessary to destroy, remove, or inhibit,
satellites
on planar orbits with large apocenters.

\section{Summary}

We are searching for a signature of hierarchical galaxy formation
in the properties of current spiral galaxies and their satellites.
Satellites at large radii, or at least the components that
would have formed those satellites, appear to have been 
preferentially ``removed'' from low inclination orbits (those in 
the disk plane) leading to
the current preferentially polar distribution of satellites (ZSFW).
Our quantitative estimate of the orbital inclination limit for
the current satellite population 
has a large uncertainty --- but, the best fit models imply
that satellites on orbits within 70$^\circ$ to 80$^\circ$ from the disk plane 
at projected radii $>$ 200 kpc have been destroyed, accreted, removed,
or 
inhibited.
We use these limits on the inclination of surviving orbits
to estimate the number of ``missing'' satellites in low inclination orbits.
The lost luminosity (or mass for constant M/L among satellites and
primaries) is consistent both with an extrapolation of the local
accretion rate
and with the hypothesis that these ``missing''
satellites contributed substantial luminosity 
($\gtsim$ 20\%) to the central galaxy.
The large statistical 
uncertainties preclude us from determining whether the
material in the ``missing'' disk plane
satellites makes a modest ($\sim$ 10\%) or dominant ($>$50\%)
contribution  to the luminosity and mass of the central galaxy.
The identification of a lost satellite population may also
help reconcile recent numerical simulations (K99, M99)
that produce many more satellites per primary than 
observed. The principal weakness of this entire discussion
is that no mechanism is demonstrated to appropriately affect
satellites with large apocenter and low orbital 
inclination relative to the primary disk.
The distribution of satellite galaxies provides a tool that,  with
more sophisticated simulations and larger samples, 
may enable us
to further develop our understanding of galaxy formation
and the dynamical evolution of galactic halos.

\vskip 1in
\noindent
\acknowledgements

DZ acknowledges partial financial support from an
NSF grant (AST-9619576), a NASA 
LTSA grant (NAG-5-3501), a David and Lucile Packard Foundation
Fellowship, and a Sloan Fellowship. 
AHG acknowledges support from an NSF Graduate Student
Fellowship. We thank A. Zabludoff for comments on a preliminary draft.

\vfill\eject

\centerline{Figure Caption}

\noindent
\figcaption{A histogram of the satellite position angle relative to 
the disk major axis. The ZSFW
sample (solid line) is compared to the predicted distribution (dashed
line) from
the Quinn \& Goodman (1986) model (adopting a limiting orbital inclination
of 45$^\circ$ and a satellite number density profile proportional to
$r^{-1.8}$). Also shown is the angular distribution of ZSFW satellites
beyond a projected radius of 200 kpc (dotted line).
}

\noindent
\figcaption{The azimuthal distribution of satellite galaxies. 
The upper panel displays the average position angle with respect to the
disk for satellites in nearly equally populated radial bins (filled
circles; left-hand axis) and the fraction of satellites to with position angles greater
than 45\degrees\ in those same bins (open circles; right-hand axis). 
The bottom panel
displays projected radius vs. position angle for the entire satellite sample.
In both panels, the bold solid line represents the results from our
simulations for Sample 1 and the best-fit $\theta_l$, 52$^\circ$.
}
\clearpage
\
\vskip 10cm
\includegraphics{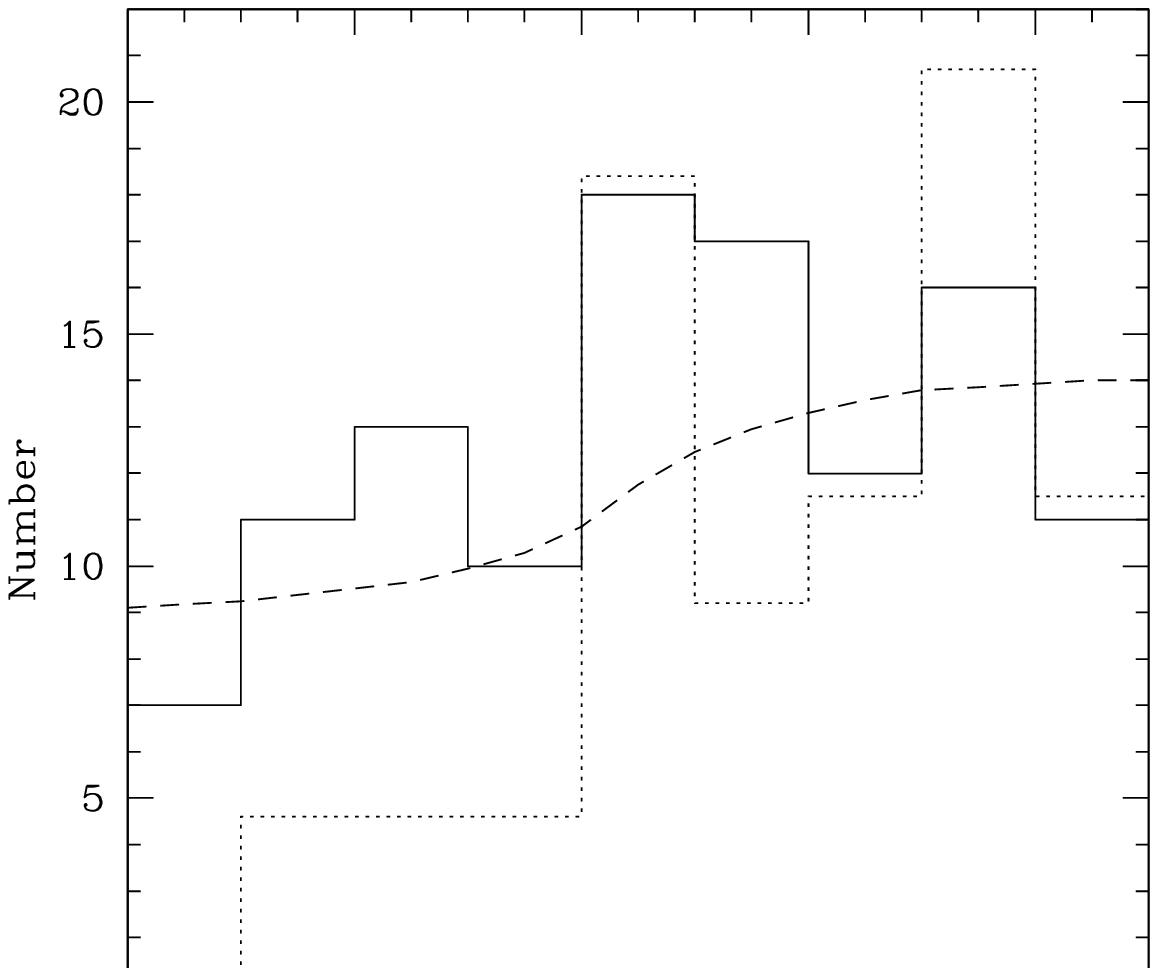}
\vskip 2cm

\clearpage
\
\vskip 14cm
\includegraphics{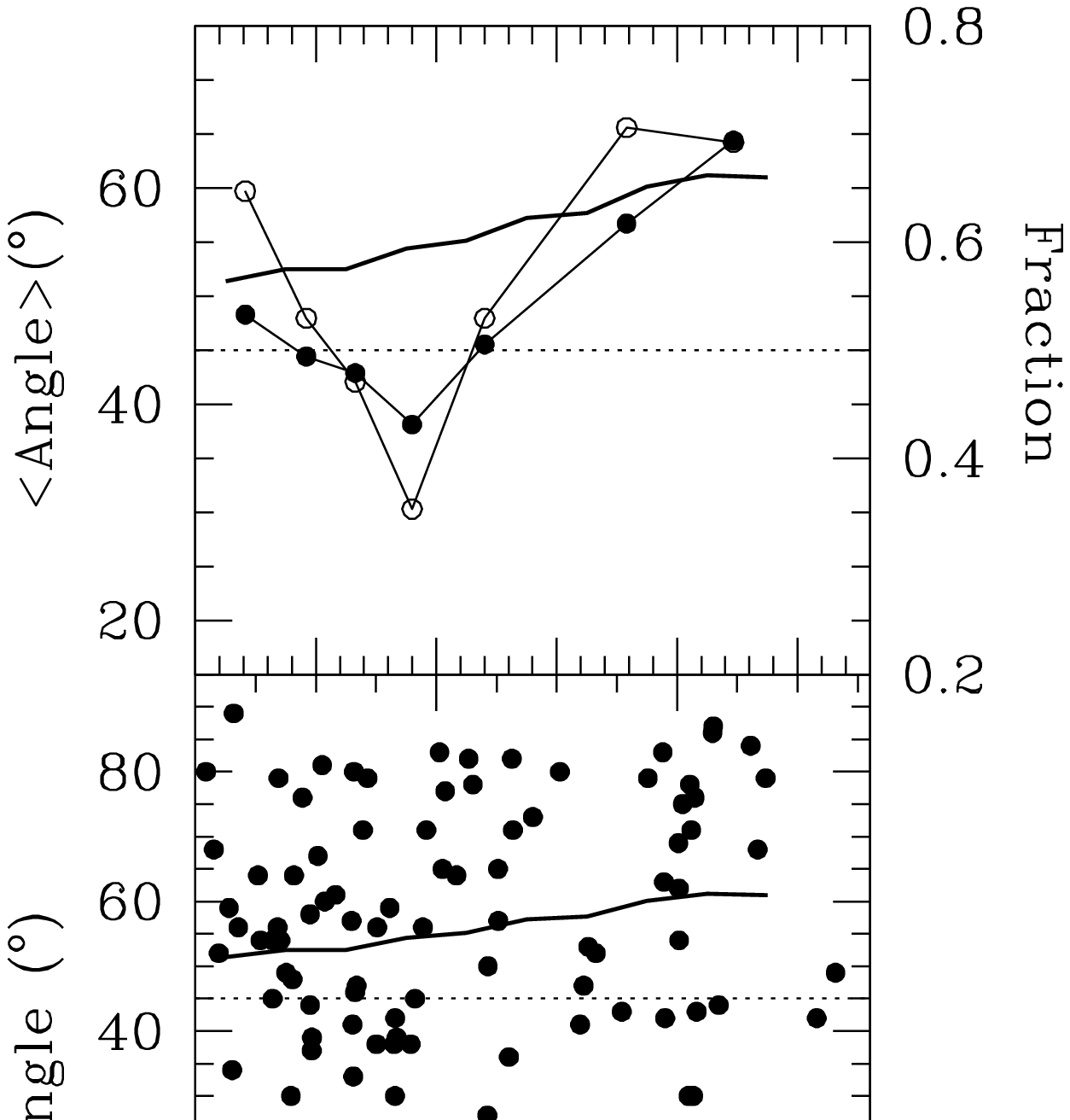}

\clearpage
\begin{deluxetable}{lcr}
\tablewidth{29pc}
\tablecaption{Satellite Samples}
\tablehead{
\colhead{Satellite Sample}             & \colhead{Orbital Inclination Limit} 
& \colhead{Proj. Radius Limit}}
\startdata
1&none&none\nl
2&none&$r_p > 300$ kpc\nl
3&$< 45^\circ$&none\nl
4&$< 45^\circ$&$r_p > 200$ kpc\nl
\enddata
\end{deluxetable}

\vfill\eject

\begin{deluxetable}{lcrrrrr}
\tablewidth{29pc}
\tablecaption{Orbit Simulations}
\tablehead{
\colhead{No.}             & \colhead{Sample No.} 
& \colhead{$\alpha$} &
\colhead{$f_{INT}$}            &\colhead{$e$} &
\colhead{$\theta_{l}$ (Kepler)} &\colhead{$\theta_{l}$ (Infall)}
\\[.2ex]
\colhead{(1)} 		   &\colhead{(2)} &\colhead{(3)}&
\colhead{(4)}		   &\colhead{(5)} &
\colhead{(6)}              &\colhead{(7)}
}
\startdata
1&4&1.8&0.10&0.5&18$^\circ$[0,48]&20$^\circ$[---,---]\nl
2&4&1.8&0.10&0.7&18$^\circ$[0,58]&12$^\circ$[---,---]\nl
3&4&1.8&0.10&0.9&32$^\circ$[10,64]&12$^\circ$[---,---]\nl
4&4&1.5&0.10&0.5&18$^\circ$[0,52]&...\nl
5&4&1.5&0.10&0.7&18$^\circ$[0,60]&...\nl
6&4&1.5&0.10&0.9&30$^\circ$[6,66]&...\nl
7&4&2.0&0.10&0.5&18$^\circ$[0,44]&...\nl
8&4&2.0&0.10&0.7&18$^\circ$[0,58]&...\nl
9&4&2.0&0.10&0.9&30$^\circ$[6,64]&...\nl
10&4&1.8&0.05&0.5&16$^\circ$[0,50]&...\nl
11&4&1.8&0.05&0.7&20$^\circ$[0,60]&...\nl
12&4&1.8&0.05&0.9&32$^\circ$[10,64]&...\nl
13&4&1.8&0.15&0.5&20$^\circ$[0,48]&...\nl
14&4&1.8&0.15&0.7&18$^\circ$[0,58]&...\nl
15&4&1.8&0.15&0.9&28$^\circ$[6,64]&...\nl
16&3&1.8&0.10&0.5&48$^\circ$[2,90]&36$^\circ$[0,90]\nl
17&3&1.8&0.10&0.7&52$^\circ$[22,90]&40$^\circ$[0,90]\nl
18&3&1.8&0.10&0.9&68$^\circ$[34,90]&40$^\circ$[0,90]\nl
19&1&1.8&0.10&0.5&38$^\circ$[14,90]&44$^\circ$[0,90]\nl
20&1&1.8&0.10&0.7&52$^\circ$[24,90]&40$^\circ$[0,90]\nl
21&1&1.8&0.10&0.9&58$^\circ$[34,90]&44$^\circ$[0,90]\nl
22&2&1.8&0.10&0.5&12$^\circ$[0,44]&12$^\circ$[---,---]\nl
23&2&1.8&0.10&0.7&26$^\circ$[0,48]&16$^\circ$[---,---]\nl
24&2&1.8&0.10&0.9&34$^\circ$[16,54]&16$^\circ$[---,---]\nl
\enddata
\end{deluxetable}
\end{document}